\documentclass[runningheads]{llncs}
\usepackage{graphicx}
\usepackage{amsmath}
\usepackage{amssymb}
\usepackage{mathabx}
\usepackage{subcaption}
\usepackage{enumitem}
\setlist[itemize]{label=$\cdot$}
\usepackage[colorlinks,citecolor=blue,urlcolor=blue, linkcolor=blue,hypertexnames=True]{hyperref}

\begin{document}
\title{ Cross-Task Data Augmentation by Pseudo-label Generation for Region Based Coronary Artery Instance Segmentation }
\titlerunning{Cross-Task Data Augmentation by Pseudo-label Generation}

\author{Sandesh Pokhrel\thanks{Equal contribution}\inst{1}\, 
Sanjay Bhandari\textsuperscript{*} \inst{1} \and
Eduard Vazquez \inst{2} \and
Yash Raj Shrestha \inst{3} \and 
Binod Bhattarai \inst{1,4}}
%
\authorrunning{Pokhrel \& Bhandari et al.}

\institute{Nepal Applied Mathematics and Informatics Institute for research(NAAMII), Lalitpur, Nepal \and
Fogsphere(Redev AI Ltd.), 64 Southwark Bridge Rd, SE1 0AS, London, UK \and
University of Lausanne, Switzerland \and
School of Natural and Computing Sciences, University of Aberdeen, Aberdeen, UK\\
\email{\{binod.bhattarai\}@abdn.ac.uk}
}
\maketitle              
\begin{abstract}
Coronary Artery Diseases (CADs) although preventable, are one of the leading causes of death and disability. Diagnosis of these diseases is often difficult and resource intensive. Angiographic imaging segmentation of the arteries has evolved as a tool of assistance that helps clinicians make an accurate diagnosis. However, due to the limited amount of data and the difficulty in curating a dataset, the task of segmentation has proven challenging. In this study, we introduce the use of pseudo-labels to address the issue of limited data in the angiographic dataset to enhance the performance of the baseline YOLO model. Unlike existing data augmentation techniques that improve the model constrained to a fixed dataset, we introduce the use of pseudo-labels generated on a dataset of separate related task to diversify and improve model performance. This method increases the baseline F1 score by 9\% in the validation data set and by 3\% in the test data set.

\end{abstract}

\section{Introduction}
The buildup of atherosclerotic plaque in the coronary arteries causes a medical condition known as Coronary Artery Disease(CAD). This obstruction of blood flow to the heart causes the arteries to clog or burst, leading to death or disability. Although CAD is preventable, it is the third leading cause of death and disability worldwide and is associated with 17.8 million deaths annually~\cite{riskfactor}. 
Physicians can determine the extent of the blockage by coronary angiography, the "gold standard" approach to the diagnosis of CAD~\cite{goldstandard}. It involves the application of a contrast agent in the arterial region to capture X-ray images of the coronary arteries. The physicians after analyzing can recommend the proper follow-up procedures, which could include revascularization of abnormal sections in the arteries. The experience of physicians influences this method of analysis of angiographic images and videos, and the diagnosis can lack accuracy, objectivity, and consistency~\cite{accuracy}. Arterial segmentation and detection of stenosis can achieve better consistency and improve accuracy of existing methods through automated procedures.

The forefront of research in invasive X-ray angiography has been through deep learning and neural networks. Automatic coronary artery detection and segmentation have been carried on using different methods like 
U-nets~\cite{localcontexttransformer,autostendiag} and  %
DenseNet~\cite{densenet}.
Following this trend we propose the use of YOLO-v8~\cite{yolov8} model for the task of segmentation of the coronary artery as the baseline. The use of this specific architecture, i.e. v8, was motivated by the fact that the aggregation of features and the Mish activation function in this model provide improved precision compared to its predecessors~\cite{Qureshi2023}. As these models require a large amount of training data to train the model, collecting such a large amount of annotated data in the medical domain in general is very challenging. We address this problem by generating pseudo-labels on a dataset curated for a separate task (stenotic region segmentation) to improve model performance of the model on target task (coronary artery segmentation). Pseudo-labels are generated on the images belonging to stenosis detection dataset through inference on YOLOv8 model which is trained on the original dataset with soft augmentations, CLAHE, Contrast-Limited Adaptive Histogram Equalisation and median blur as these augmentations enhance x-ray images and create better separability \cite{CLAHEenhancement}. 

Pseudo-labels, unlike traditional data augmentation techniques such as different geometric transformations, not only increase the size of the dataset but also introduce greater diversity in training samples and help solve the problem of data sparsity in medical imaging by encouraging the utilization of unlabelled datasets of similar nature. We further compare this with fully self supervised approach MaskDino~\cite{maskdino} and ConvNeXt~\cite{convnext}, ConvNextV2~\cite{Convnextv2} based Mask R-CNN~\cite{maskrcnn} methods, all of which are outperformed by the proposed YOLOv8Pse model.

\section{Literature Review}

There are two main streams that diagnose CADs, invasive and non-invasive. Non-invasive methods~\cite{noninvasiveECG,SPECT-MPI} although promising fail to deliver the same effectiveness as the "gold standard"~\cite{goldstandard} invasive methods of treatment.
Automation in detection and diagnosis of CADs using non-invasive deep learning methods include analysis of ECG~\cite{noninvasiveECG,ecgstenosis} or SPECT-MPI~\cite{SPECT-MPI} signals to derive distinctive features that relate to a normal person's heart and one with CADs. 
In automation of invasive angiography images, methods like automatic segmentation of arteries based on multi scale Gabor and Gaussian filters along with multi layer perceptrons ~\cite{automaticMultiScale} and automatically identifying and classifying the angle of view and then identifying region of interest of stenosis ~\cite{automatedsten} have been explored. Further, utilizing the effectiveness of Mask R-CNN~\cite{maskrcnn} in medical domain, Fu et.al.~\cite{maskrcnnCCTA} proposed using it for segmentation which showed promising results on fine and tubular structures of the coronary arteries. U-nets~\cite{unet-orig} explored widely in medical image segmentation have several modifications that work well in angiographic images. 

Newer architectures such as ConvNeXt have been explored in medical imaging. BCU-net~\cite{bcunet} used ConvNeXt~\cite{convnext} in global interaction and U-Net~\cite{unet-orig} in local processing in binary classification.
Specifically in tasks related to arterial segmentation, ConvNeXt has been used to improve the classification of RCA angiograms using LCA information~\cite{convnextangiogram}. The latest iteration~\cite{Convnextv2} that has improved performance over ConvNeXt due to architectural improvements is much less explored in medical imaging tasks and even less so in angiographic images. 
In angiographic setting ConvNeXtv2 \cite{convnxtv2medicalimaging} has shown it's effectiveness in instance segmentation. YOLO models~\cite{yoloorig} have been emerging in medical image analysis due to their real-time inference capability and versatility in object detection. A much more comprehensive analysis of various YOLO algorithms that were explored in medical imaging from 2018 shows the trend for the improved versions in their ability to extract features as well as in downstream tasks due to their specialised heads~\cite{Qureshi2023}. Identifying images for specific tasks is a bottleneck in many deep learning problems. This problem is more pronounced in medical imaging, especially due to the lack of expert annotators in the relevant domains. To overcome similar problems in other domains, pseudo-labels have been seen as a strong form of data augmentation. Self-supervised learning approaches, effectively utilise unlabelled dataset to improve the model performance, usually associated with varied form of data augmentation~\cite{Pseudolabelsemseg,pseudolabelsimple,pesudolabelgeneration}. 

Inspired by the trend of YOLO in medical imaging and the effectiveness of pseudo-labels, we propose the use of the YOLOv8 architecture with to generate on angiographic stenosis images to aid the segmentation of coronary arteries which we term 'Yolov8Pse'. We compare our pipeline method to MaskDino, ConvNeXt as well as ConvNeXtv2 and show that our pipeline in YOLO is stronger than these baseline methods in instance segmentation of multiple classes.

\section{Methods}
We study the impact of pseudo-labeling on the performance of instance segmentation models, specifically in the context of angiographic images. Our method focuses on how pseudo-labels on unlabeled dataset can mitigate the limitations imposed by a small number of labeled data points, thereby enhancing the robustness and accuracy of instance segmentation models. By leveraging pseudo-labeling, we aim to improve the model's ability to generalize from limited labeled data and achieve superior performance in segmenting angiographic images.

\subsection{Problem Formulation}
We focus on supervised instance segmentation for coronary artery segmentation, where \(X\) represents the input space, and \(Y\) denotes the segmentation mask and label space for segmentation classes. Let \(\text{D}_{\text{artery}} = \{(x_i, y_i)\}_{i=1}^{n_{\text{artery}}}\) be the original artery segmentation dataset, which is i.i.d. from \(P_{X, Y_{\text{artery}}}\), and \(\text{D}_{\text{stenosis}} = \{(x_i)\}_{i=1}^{n_{\text{stenosis}}}\) be the stenosis detection dataset which is pseudolabeled to generate pseudolabeled stenosis dataset \(\text{D}_{\text{pseudo-stenosis}} = \{(x_i, \hat{y}_i)\}_{i=1}^{n_{\text{stenosis}}}\) . We propose to use an anchor-free YOLOv8 model for artery segmentation task, a model consisting of CSP53 Darknet as backbone, which consists of 53 convolutional layers and employs cross-stage partial connections to improve information flow between the different layers.

\begin{figure}
\includegraphics[width=\linewidth, scale=1]{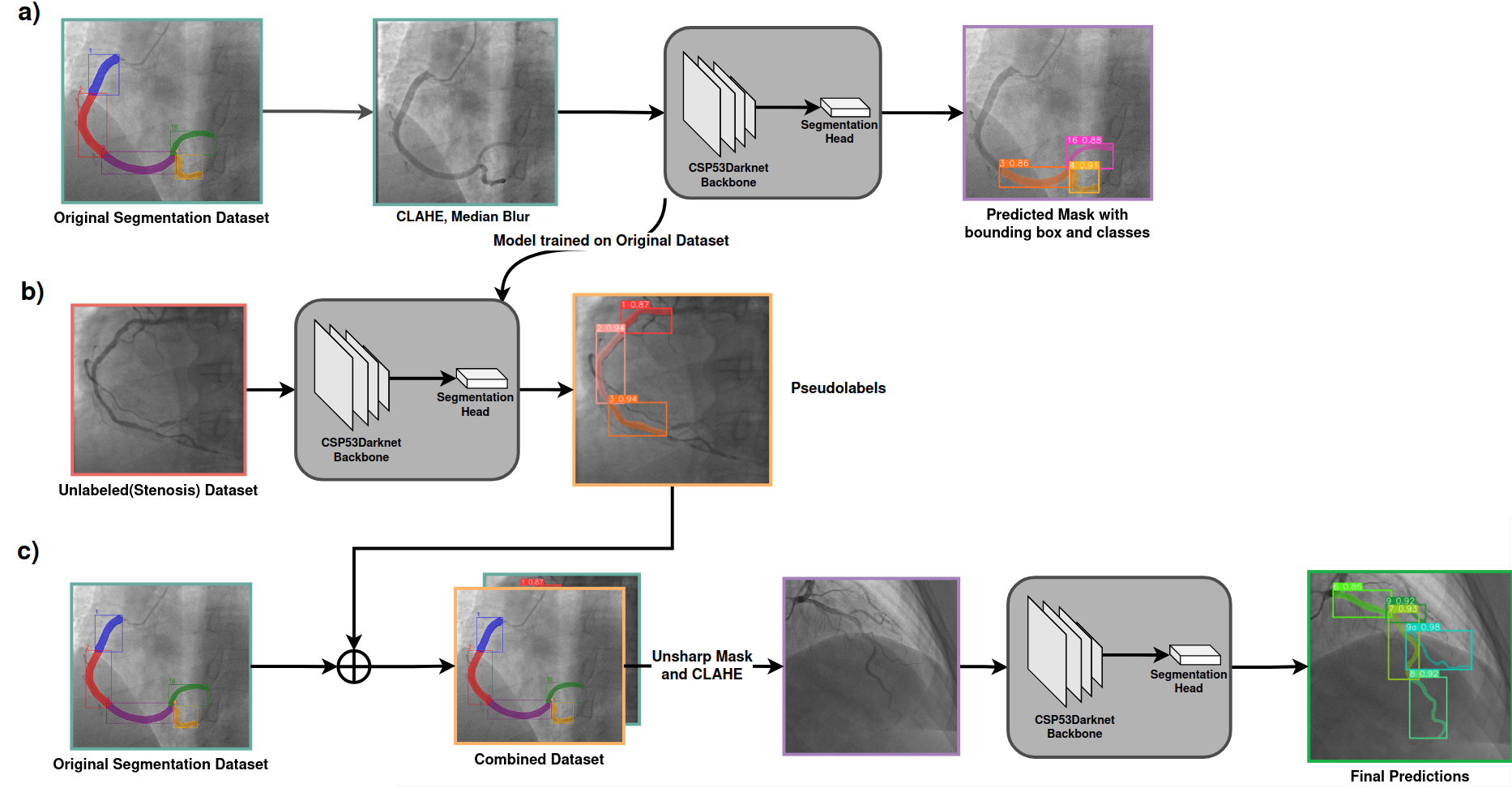}
\caption{a) Training of the intermediate model on the original artery segmentation dataset b) Generating pseudo-labels for the unlabeled stenosis detection dataset c) Training the model on a combination of the original artery segmentation dataset and the resultant pseudo-labeled dataset. } \label{fig1}
\end{figure}

\subsection{Cross-Task Dataset Pseudo-labelling} 
The pseudo-label generation procedure is presented in \ref{fig1}[a), b)]. A soft composite augmentation(CLAHE and median blur) is done on the original images before training an instance segmentation model. CLAHE, Contrast-Limited Adaptive Histogram Equalisation, enhances the features of x-ray angiographic images~\cite{CLAHEenhancement}, while median blur mollifies any sharp artefacts that might be added due to CLAHE. The intermediate instance segmentation model \(f: X \rightarrow \mathbb{R}^{|Y|}\) is trained with these augmented images from \(\text{D}_{\text{artery}}\). As shown in Figure \ref{fig1}.b, the trained model \(f\) generates pseudolabels \(\hat{y}_i\) for the stenosis detection dataset by predicting segmentation masks, i.e., for \(x_i \in \text{D}_{\text{stenosis}}\), \(\hat{y}_i = f(x_i)\), thereby creating the pseudolabeled dataset \(\text{D}_{\text{pseudo-stenosis}} = \{(x_i, \hat{y}_i)\}_{i=1}^{n_{\text{stenosis}}}\).  The combined dataset \(\text{D}_{\text{combined}}\) is constructed by merging \(\text{D}_{\text{artery}}\) and \(\text{D}_{\text{pseudo-stenosis}}\), i.e.,
\begin{equation}
\text{D}_{\text{combined}} = \text{D}_{\text{artery}} \cup \text{D}_{\text{pseudo-stenosis}}
\label{combinationeqn}
\end{equation}

The input to the final YOLOv8 model(Yolov8Pse) is a combined dataset  \(\text{D}_{\text{combined}}\) consisting of images and annotation from the segmentation dataset, as well as the images and pseudo-labels generated on the stenosis dataset using the previously trained model. Following this the combined dataset goes through training time augmentation of unsharp mask to bring out higher level edge details of the arteries for better separability. The final model training procedure is shown in Fig1.c).\ref{fig1}

\subsection{Loss Function}
The loss function for the instance segmentation downstream task for YOLOv8 \cite{yolov8} is the sum of IOU loss, mask loss, class loss, and Distributional focal loss with different gain coefficients adjusted to obtain better performance.
\begin{equation}
\mathcal{L} = \lambda_c . \mathcal{L}_c +  \lambda_f . \mathcal{L}_f + \lambda_s . \mathcal{L}_s + \lambda_b . \mathcal{L}_b
\label{losseqn}
\end{equation}

where, $\mathcal{L}_c$ is the multilabel classification BCE, $\mathcal{L}_f$ represents the distributional focal loss, $\mathcal{L}_s$ is the BCE loss for segmentation and $\mathcal{L}_b$ is the IOU-Loss. The corresponding $\lambda$ values are gain coefficients for each loss function treated as hyperparameters. We further explore the values of hyperparameters and  their choice in implementation section.

The loss function \ref{losseqn} is optimized on $D_{artery}$ in the first phase. This model is used to predict labels on $D_{stenosis}$ dataset and curate the pseudo-labels dataset, $D_{pseudo-stenosis}$. Finally, following \ref{combinationeqn}, we obtain the new dataset on which the same YOLOv8 instance segmentation loss function \ref{losseqn} is minimized to obtain the final model trained on the $D_{combined}$ dataset.

\section{Experiments}
\subsection{Dataset and Preprocessing}
The ARCADE dataset~\cite{Arcadedataset} consists of 1200 images each for coronary artery segmentation and stenosis detection. The spatial size of the images in dataset is 512 x 512 px. The pseudo-labels were generated on the stenosis dataset from a YOLO-v8 instance segmentation model trained on the segmentation dataset with soft augmentations(CLAHE and Median Blur). Predictions with confidence score 0.5 or greater were considered valid annotations and saved for a new training dataset consisting of pseudo-labels from stenosis dataset and the original segmentation dataset. The new combined dataset, including the validation set, was pre-processed using a series of image enhancement techniques. These techniques included the application of an Unsharp Mask filter followed by  Contrast Limited Adaptive Histogram Equalization (CLAHE). This enhancement was applied to effectively improve the quality of the dataset for further analysis and model training.

\subsection{Implementation details}
In this study, we use the YOLO-v8 segmentation model for vessel segmentation following a augmented pseudo-labeling pipeline. YOLO-v8 segmentation model was trained on NVIDIA RTX 3090 graphic card for 120 epochs using AdamW optimizer ($\beta1= 0.9$, $\beta2 = 0.999$) with an initial learning rate of 1 × $10^{-2}$ and a decay rate of 0.0005 per epoch with batch size of 16. And a series of random augmentations were applied to the dataset to increase the diversity in training examples which included HSV Hue Adjustment, HSV Saturation Adjustment,HSV Value Adjustment, Translation, Scaling,Vertical Flipping and Horizontal Flipping. The coefficients for gains of loss functions we used are $\lambda_b=7.5, \lambda_c= 0.5, \lambda_s = 0.468, \lambda_f = 2.0$. We settled on these values after carefully evaluating the performance of our model at different settings. The selected gains allowed the model achieve smoother training.

\subsection{Quantitative Results}
The quantitative result for the segmentation of the coronary arteries with 25 classes is compared in Table \ref{tab:Vessel segmentation}. The F1 scores for corresponding models on the validation set as well as the test set are compared as metrics.The results presented in Table \ref{tab:Vessel segmentation} demonstrate that incorporating pseudo-labels into the YOLOv8 model yields significantly better performance compared to an ensemble of differently initialized YOLOv8 models. 

Specifically, the F1 score improved from 0.34 to 0.41 on the validation dataset and from 0.26 to 0.35 on the test dataset when comparing the pseudo-labeled YOLOv8 to the YOLOv8 ensemble. Moreover, when comparing the pseudo-labeled YOLOv8 directly to the standalone YOLOv8 model, a notable performance increase of 0.9 F1 score on the validation dataset and 0.2 F1 score on the test dataset is observed. This highlights the importance and strength of pseudolabeling in enhancing model performance by effectively leveraging unlabeled data to improve predictive accuracy
\begin{table}[t]
    \centering
    \begin{tabular}{c c c} \hline
         \textbf{Architecture}&  \textbf{F1Score(Val)$\uparrow$}& \textbf{F1Score(Test)$\uparrow$}\\ \hline 
         Yolov8 Ensemble& 0.34 & 0.26 \\ 
         ConvNext~\cite{convnext}
         & 0.21 & 0.27\\ 
         ConvNextv2~\cite{Convnextv2}
         & 0.26 & 0.29\\ 
         MaskDino~\cite{maskdino}&  0.38 & 0.33\\ 
         \hline 
         Yolov8~\cite{yolov8}&  0.32 & 0.33\\ 
         Yolov8Pse (Ours)&  \textbf{0.41} & \textbf{0.35}\\ \hline
    \end{tabular}\\
    \captionsetup{skip=5pt}
    \caption{Comparison of F1 scores of different architectures on Arcade coronary artery segmentation dataset}
    \label{tab:Vessel segmentation}
\end{table}

\begin{table}
    \centering
    \begin{tabular}{c c} \hline
    \textbf{Architecture}& \textbf{mAP/50(Val)}$\uparrow$\\ \hline
         MaskDino& 0.54\\ 
         Yolov8&  0.53\\ 
         ConvNeXtv2& 0.46 \\ \hline
         Yolov8Pse (Ours)& \textbf{0.59}\\ \hline
    \end{tabular}
    \captionsetup{skip=5pt}
    \caption{mAP/50 scores achieved on the validation set}
    \label{tab:vessel segmentation map scores}
\end{table}
The quality of detection was also be tracked through mAP scores obtained by the models on the validation set to verify that they will have generalizability in the test set. The following table shows the best mAP/50 scores achieved by the corresponding models in the validation set.

\subsection{Qualitative Results}

Qualitative comparison of the baseline models along with ours on the unseen test dataset indicates that the proposed model showcases finer attention to detail compared to other SOTA instance segmentation methods. For images with good contrast between the vessel and the background, almost all of the baseline models perform well in detection as well as segmentation. However, when the contrast and illumination changes occur in input images, baseline models make seem more erratic and make faulty predictions or are missing predictions altogether.

\begin{figure}[t]
    \centering
    \begin{subfigure}[b]{0.15\textwidth}
        \centering
        \caption*{GT}
        \includegraphics[width=2cm,height=2cm]{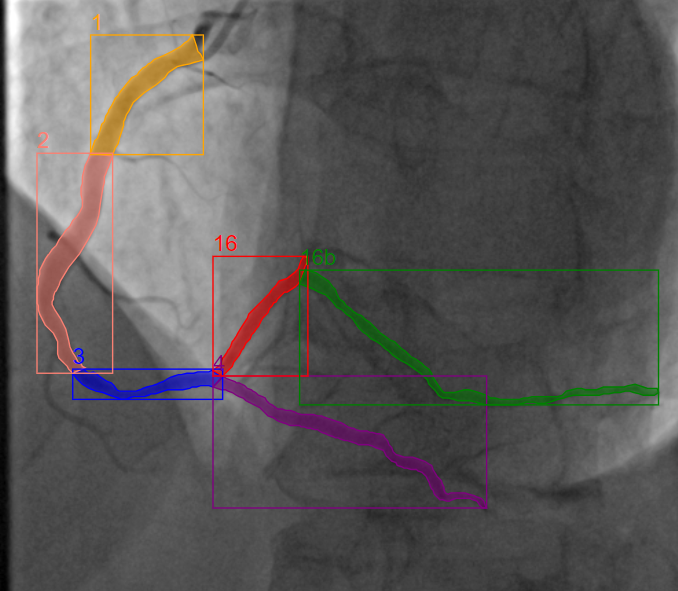}
        \label{fig:gt}
    \end{subfigure}
    \hspace{0.1cm}
    \begin{subfigure}[b]{0.15\textwidth}
        \centering
        \caption*{ConvnextV2}
        \includegraphics[width=2cm,height=2cm]{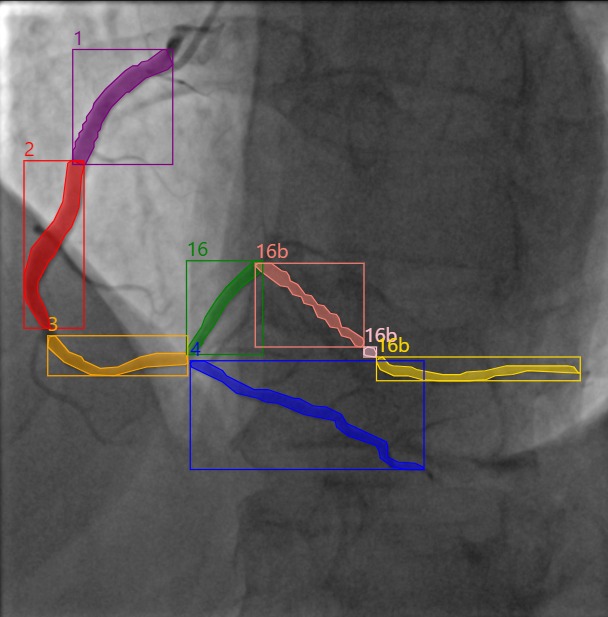}
        \label{fig:convnxtv2}
    \end{subfigure}
    \hspace{0.1cm}
    \begin{subfigure}[b]{0.15\textwidth}
        \centering
        \caption*{Yolov8}
        \includegraphics[width=2cm,height=2cm]{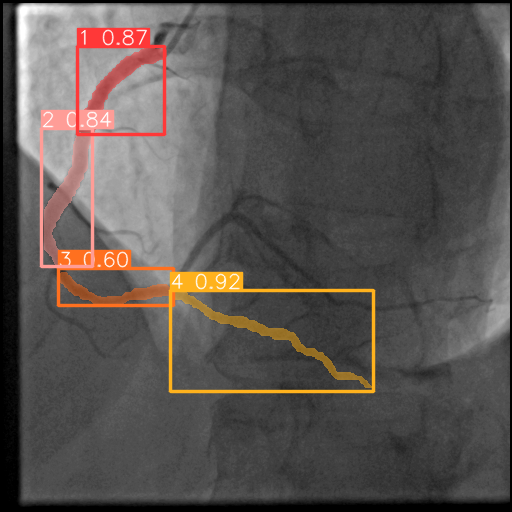}
        \label{fig:yolo2}
    \end{subfigure}
    \hspace{0.1cm}
    \begin{subfigure}[b]{0.15\textwidth}
        \centering
        \caption*{MaskDino}
        \includegraphics[width=2cm,height=2cm]{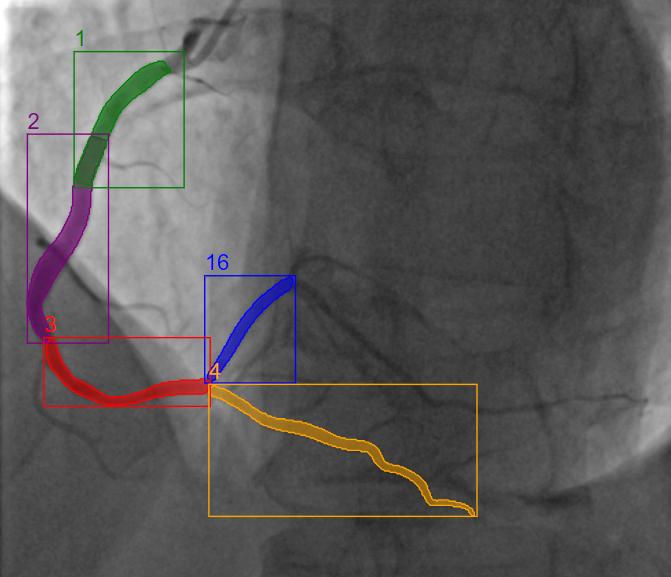}
        \label{fig:maskdino2}
    \end{subfigure}
    \hspace{0.1cm}
    \begin{subfigure}[b]{0.15\textwidth}
        \centering
        \caption*{Yolov8Pse}
        \includegraphics[width=2cm,height=2cm]{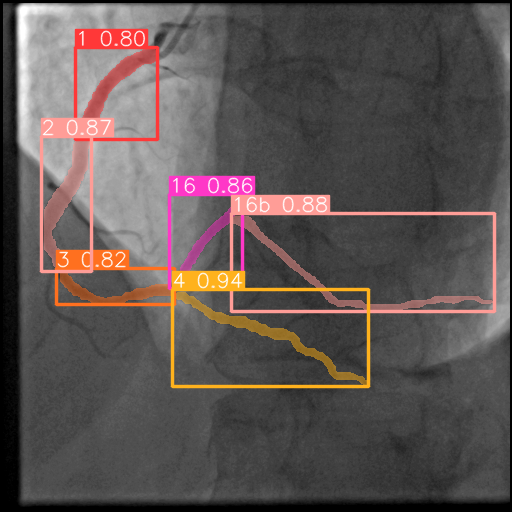}
        \label{fig:pseudoyolo}
    \end{subfigure}
    \vspace{0.1cm}
    \begin{subfigure}[b]{0.15\textwidth}
        \centering
        \includegraphics[width=2cm,height=2cm]{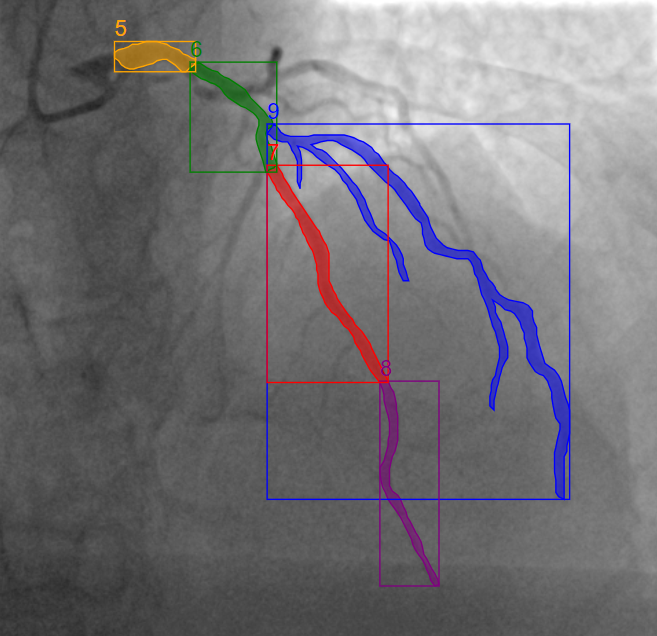}
        \label{fig:gt1}
    \end{subfigure}
    \hspace{0.1cm}
    \begin{subfigure}[b]{0.15\textwidth}
        \centering
        \includegraphics[width=2cm,height=2cm]{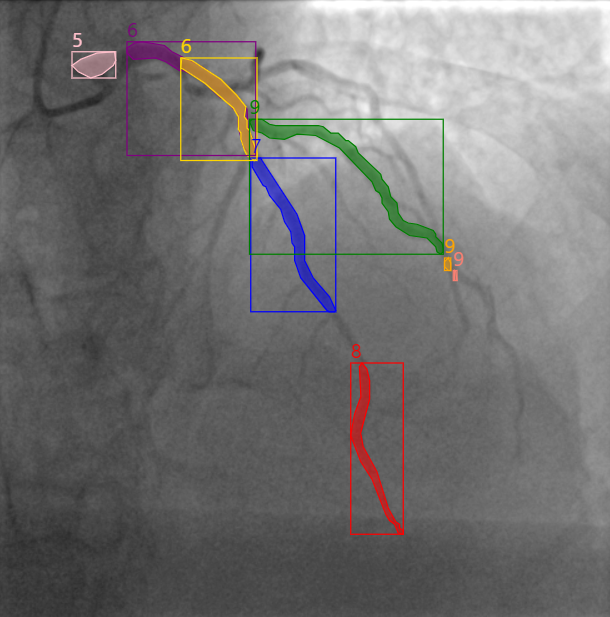}
        \label{fig:convnxtv22}
    \end{subfigure}
    \hspace{0.1cm}
    \begin{subfigure}[b]{0.15\textwidth}
        \centering
        \includegraphics[width=2cm,height=2cm]{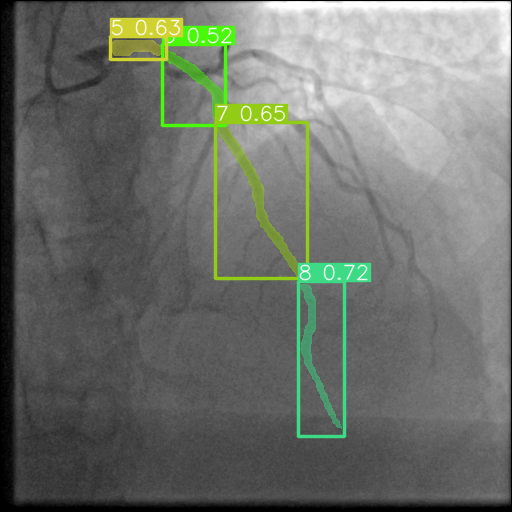}
        \label{fig:yolo3}
    \end{subfigure}
    \hspace{0.1cm}
    \begin{subfigure}[b]{0.15\textwidth}
        \centering
        \includegraphics[width=2cm,height=2cm]{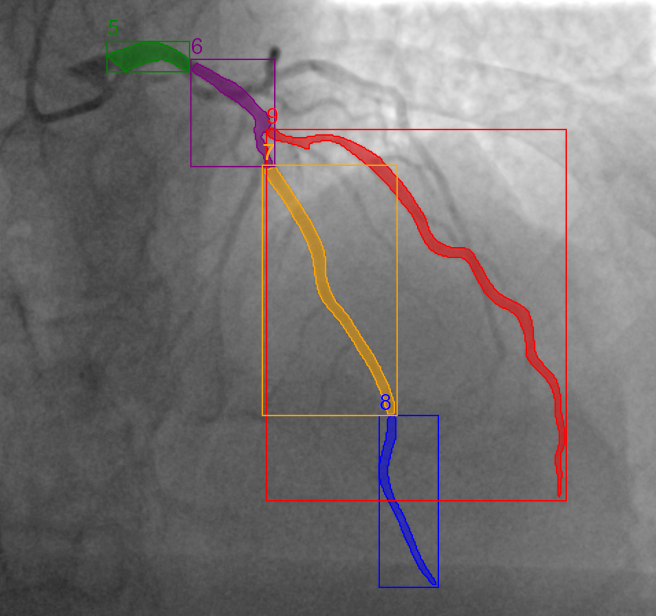}
        \label{fig:maskdino3}
    \end{subfigure}
    \hspace{0.1cm}
    \begin{subfigure}[b]{0.15\textwidth}
        \centering
        \includegraphics[width=2cm,height=2cm]{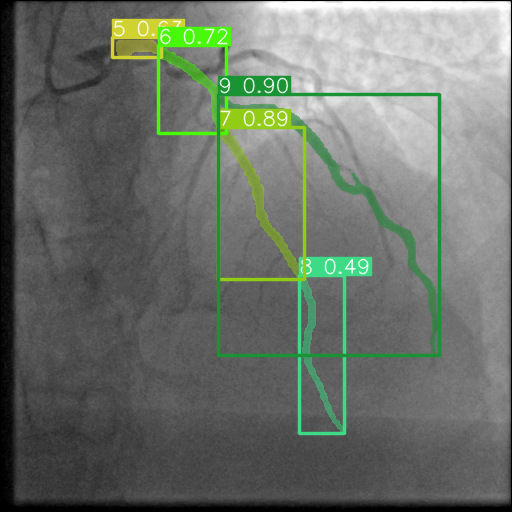}
        \label{fig:pseudoyolo1}
    \end{subfigure}

    \vspace{0.1cm}
    \begin{subfigure}[b]{0.15\textwidth}
        \centering
        \includegraphics[width=2cm,height=2cm]{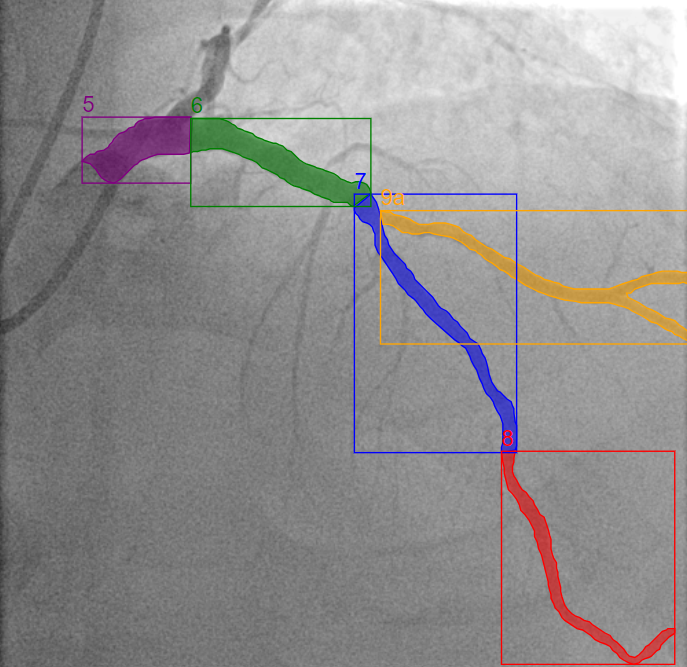}
        \label{fig:gt2}
    \end{subfigure}
    \hspace{0.1cm}
    \begin{subfigure}[b]{0.15\textwidth}
        \centering
        \includegraphics[width=2cm,height=2cm]{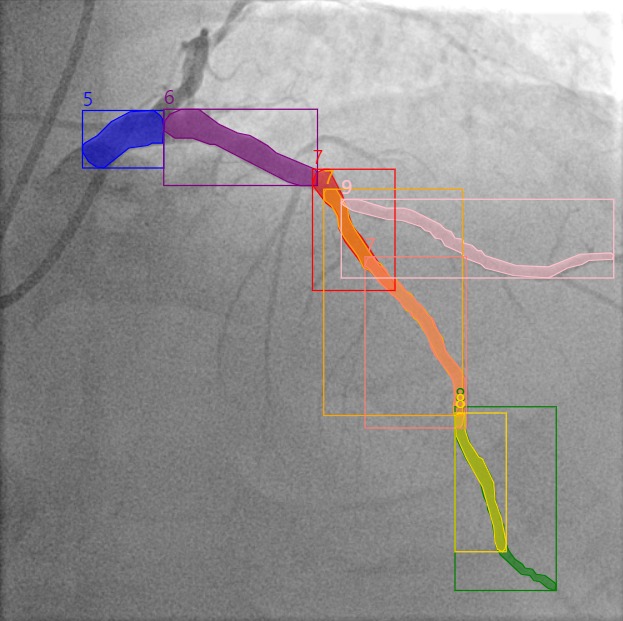}
        \label{fig:convnxtv23}
    \end{subfigure}
    \hspace{0.1cm}
    \begin{subfigure}[b]{0.15\textwidth}
        \centering
        \includegraphics[width=2cm,height=2cm]{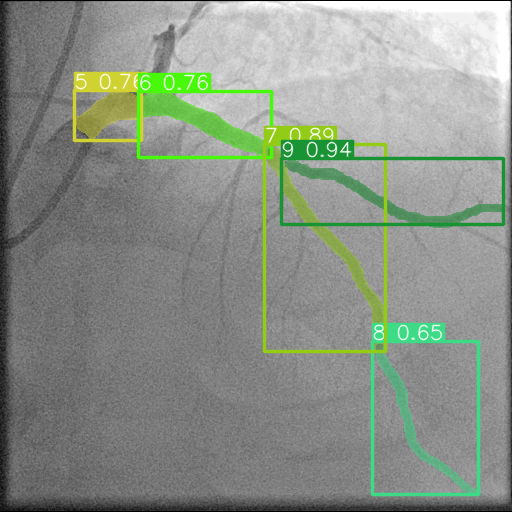}
        \label{fig:yolo4}
    \end{subfigure}
    \hspace{0.1cm}
    \begin{subfigure}[b]{0.15\textwidth}
        \centering
        \includegraphics[width=2cm,height=2cm]{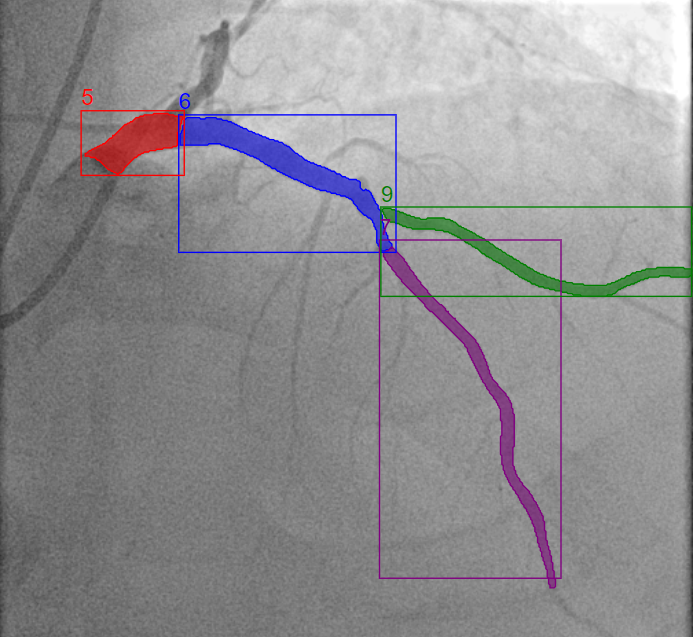}
        \label{fig:maskdino4 }
    \end{subfigure}
    \hspace{0.1cm}
    \begin{subfigure}[b]{0.15\textwidth}
        \centering
        \includegraphics[width=2cm,height=2cm]{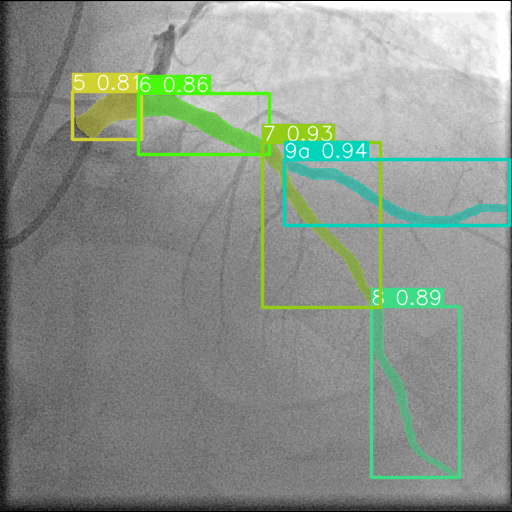}
        \label{fig:psuedoyolo1}
    \end{subfigure}

        \vspace{0.1cm}
    \begin{subfigure}[b]{0.15\textwidth}
        \centering
        \includegraphics[width=2cm,height=2cm]{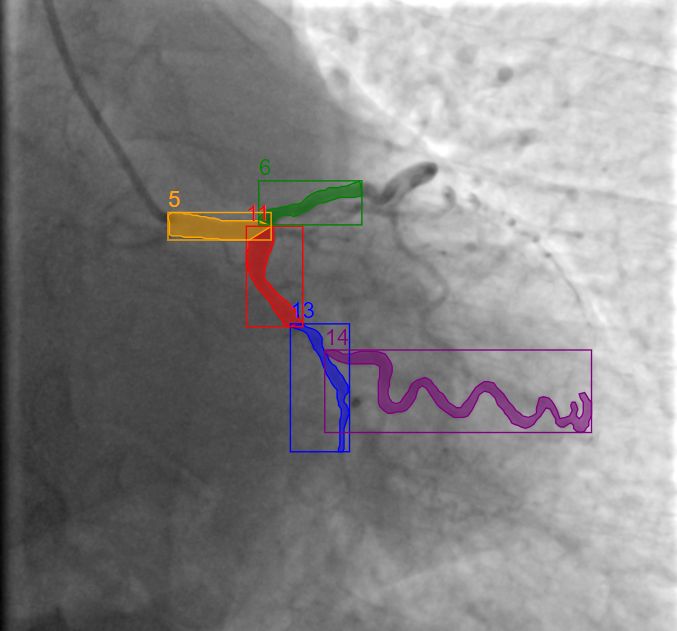}
        \label{fig:gt3}
    \end{subfigure}
    \hspace{0.1cm}
    \begin{subfigure}[b]{0.15\textwidth}
        \centering
        \includegraphics[width=2cm,height=2cm]{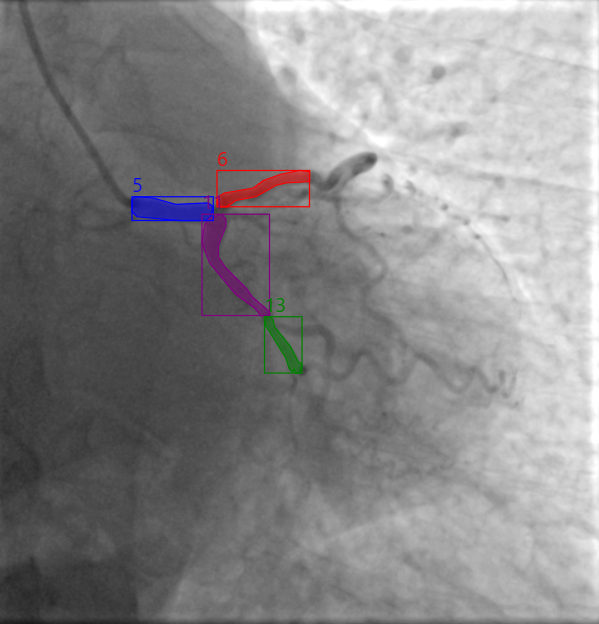}
        \label{fig:convnxtv24}
    \end{subfigure}
    \hspace{0.1cm}
    \begin{subfigure}[b]{0.15\textwidth}
        \centering
        \includegraphics[width=2cm,height=2cm]{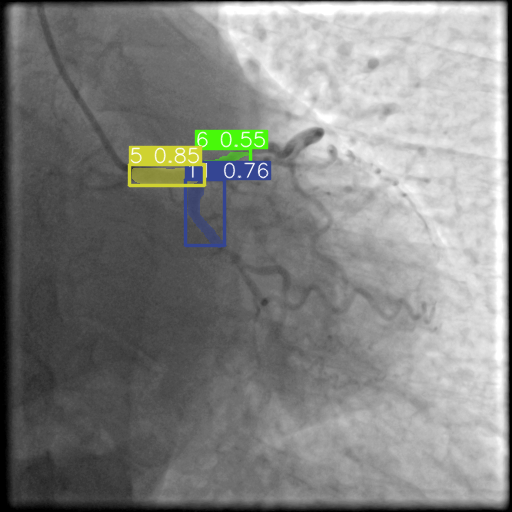}
        \label{fig:yolo1}
    \end{subfigure}
    \hspace{0.1cm}
    \begin{subfigure}[b]{0.15\textwidth}
        \centering
        \includegraphics[width=2cm,height=2cm]{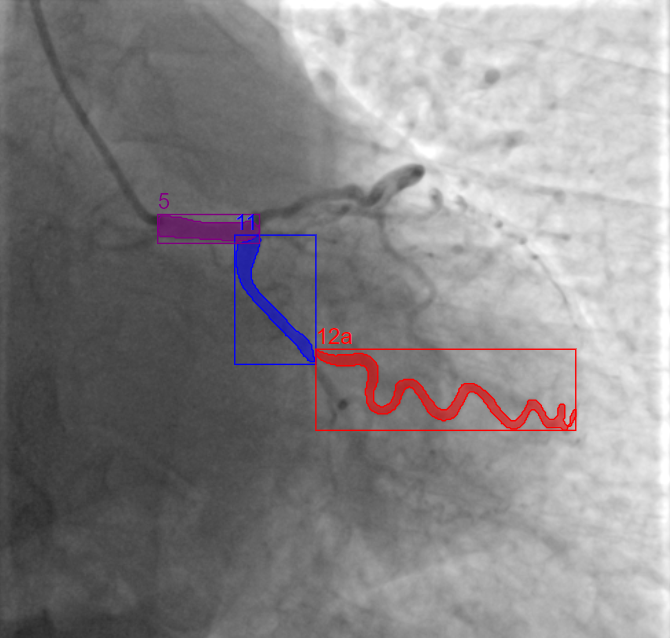}
        \label{fig:maskdino1 }
    \end{subfigure}
    \hspace{0.1cm}
    \begin{subfigure}[b]{0.15\textwidth}
        \centering
        \includegraphics[width=2cm,height=2cm]{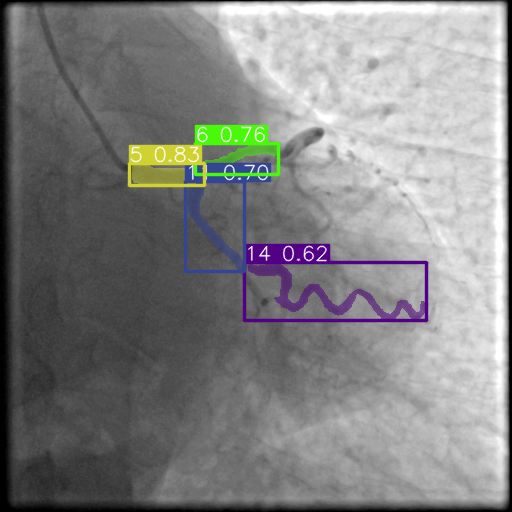}
        \label{fig:psuedoyolo2}
    \end{subfigure}
    
    \captionsetup{skip=5pt}
    \caption{Qualitative instance segmentation results on Vessel segmenation. Ground truth masks followed by the instance segmentation masks generated by ConvnextV2, Yolov8, MaskDino and Pseudolabel-Yolo(Yolov8Pse) are shown in the figure respectively.}
    \label{fig:Qualitative comparison of Segmentation}
\end{figure}

\subsection{Ablation Studies}

Starting a model from scratch was inefficient as the segmentation dataset was relatively small with only 1200 examples. Thus, the models we tested were initialized on pretrained MS-COCO dataset which have already learned a ton of essential features necessary for  segmenting objects. 

Another important improvement in our model was achieved when instead of single best model we used an average of five models exported from near the end of the training phases. We associate this improvement with the fact that the model will have learned most of the features during later epochs and a combination of these weights act as an ensemble to improve model performance. The averaged model achieved better performance on F1 Score when compared to the best model from previous training owing to the improved generalizability when averaged. Table \ref{tab:f1 score} illustrates this fact.
\begin{table}
    \centering
    \begin{tabular}{c c} \hline 
         \textbf{Model}& \textbf{F1-score}($\uparrow$)\\ \hline 
          Yolov8Pse & 0.34\\ 
         Avg. Yolov8Pse \t &\textbf{0.35}\\ \hline 
    \end{tabular}
    \captionsetup{skip=5pt}
    \caption{Comparison of weight averaged model against single best YOLOv8 model trained through the pseudo-label pipeline.}
    \label{tab:f1 score}
\end{table}

\section{Conclusion}
Our approach tackles the problem of sparsity of data and difficulty in curating a new one by adopting augmentations via pseudo-labeling on cross-task dataset to improve upon the baseline models. The importance of pseudo-labeling cross-task dataset and how we can leverage related datasets in medical scenarios has been highlighted in our work. Especially, we discuss how pseudo-labels expand the training dataset allowing the model to learn from a diverse environment and has larger range of training examples compared to traditional data augmentation approach where only geometric transformations are made to introduce diversity in a fixed data setting. 

With the extensive support of pseudo labels combined with traditional data augmentation techniques like CLAHE, median blur and unsharp mask, we improve the performance of Yolov8 in instance segmentation task, proposed as Yolov8Pse, by 9\%(F1 score). Yolov8Pse outperforms well-established state of the art instance segmentation models like MaskDino(self supervised vision transformer), ConvNeXt and ConvNeXtv2 tested with 25 classes of coronary artery segments in x-ray angiography images.

\subsubsection{Acknowledgments:} This project is supported by Leading House South Asia and Iran, Zurich University of Applied Sciences.

\bibliographystyle{splncs04.bst}
\bibliography{reference.bib}

\end{document}